\documentclass[aps,twocolumn,prb,showpacs,preprintnumbers,amsmath,amssymb]{revtex4}
\newcommand{\lsim}
 {\ \raise.35ex\hbox{$<$}\kern-0.75em\lower.5ex\hbox{$\sim$}\ }
\newcommand{\gsim}
 {\ \raise.35ex\hbox{$>$}\kern-0.75em\lower.5ex\hbox{$\sim$}\ }

\usepackage{graphicx}
\usepackage{dcolumn}
\usepackage{bm}
\usepackage{amsmath}
\usepackage{times}
\usepackage{braket}
\usepackage[usenames]{color}

\usepackage{ulem}

\begin{document}
%\draft
%\preprint{?????}
\title{Photoinduced charge-order melting dynamics in a one-dimensional interacting Holstein model}
\author{Hiroshi~Hashimoto  and Sumio~Ishihara}
 \affiliation{Department  of Physics,  Tohoku University,  Sendai 980-8578,  Japan}
%  \affiliation{CREST-JST,  Sendai 980-8578,  Japan}
\date{\today}
\begin{abstract}
Transient quantum dynamics in an interacting fermion-phonon system are investigated. In particular, a charge order (CO) melting after a short optical-pulse irradiation and roles of the quantum phonons on the transient dynamics are focused on. 
A spinless-fermion model in a one-dimensional chain coupled with local phonons is analyzed numerically. 
The infinite time-evolving block decimation algorithm is adopted as a reliable numerical method for  one-dimensional quantum many-body systems. 
Numerical results for the photoinduced CO melting dynamics without phonons are well interpreted by the soliton picture for the CO domains. 
This interpretation is confirmed by the numerical simulation for an artificial local excitation and the classical  soliton model. 
In the case of the large phonon frequency corresponding to the antiadiabatic condition, 
the CO melting is induced by propagations of the polaronic solitons with the renormalized soliton velocity. 
On the other hand, in the case of the small phonon frequency corresponding to the adiabatic condition, 
the first stage of the CO melting dynamics occurs due to the  energy transfer from the fermionic to phononic  systems, and the second stage is brought about by the soliton motions around the bottom of the soliton band.  Present analyses provide a standard reference for the photoinduced CO melting dynamics in low-dimensional  many-body quantum systems. 
\end{abstract}
%\pacs{ \red{78.47.J-, 78.20.Bh, 74.72.-h}}
%78.47.J- 	Ultrafast spectroscopy (<1 psec)
%78.20.Bh 	Theory, models, and numerical simulation
%74.72.-h 	Cuprate superconductors
\maketitle
\section{Introduction}

Ultrafast dynamics in correlated electron systems induced by the intensive optical laser pulse have been widely accepted as an attractive theme in recent solid-state physics.~\cite{nasu,PIPT_JPSJ}
This issue is motivated not only from the fundamental research in highly nonequilibrium many-body systems  but also from the potential applications to the efficient energy conversion and ultrafast optical switching.  
One of the prototypical targets for the nonequilibrium many-body systems is the Mott insulator and the carrier-doped Mott insulator. 
A variety of exotic nonequlibrium phenomena emerges owing to the spin and orbital degrees of freedom,~\cite{han,matsuda,iwai,okamoto,novelli,stojchevska,cavalleri,kubler,perfetti}
and a number of theoretical studies has been also performed in the optical induced nonequilibrium state in the Mott insulators.~\cite{takahashi,oka,freericks,moritz,meisner,werner,vidmar,mierzejewski,lu,iyoda} 

Another promising target for the photoinduced nonequilibrium dynamics in correlated electron systems is 
the charge ordered (CO) state, in which the averaged electron density per site is 0.5 in most of the cases. 
The CO states, in which the translational symmetry of the electronic charge distribution is broken, are  ubiquitously observed in a wide class of solids. 
% such as the transition-metal compounds, the organic molecular solids, and the rare-earth metal compounds.  
On the verge of CO, several rich phenomena emerge, such as superconductivity, metal-insulator transition, ferroelectricity, colossal magneto-resistance and so on. 
Since the CO system is susceptible against the external stimuli due to its small energy scale in comparison with the Mott insulator, this system is a plausible candidate for the nonequilibrium electron dynamics research.~\cite{kiryukhin,poli,beaud,iwai2,ichikawa,onda,iwano09,rincon,hashimoto14,hashimoto15} 
Actually, a variety of CO compounds have been investigated by the ultrafast optical pump-probe measurements as well as the several time-resolved experimental techniques.
In contrast to the Mott insulators, the spin degree of freedom hides behind. 
%ectron density, i.e. the quarter filling. 
Instead, the lattice degree of freedom plays crucial roles for the equilibrium and nonequilibrium states as follows:  
i) The non-uniform electronic charge distribution and the alternate lattice distortion are stabilized cooperatively. 
ii) Different energy and time scales in the electronic and phononic systems provide a rich variety of the transient dynamics.  
iii) The energy transfer possibly occurs from the highly excited electronic to phononic systems.  

During the several CO systems,  one dimensional CO insulator is a simple example for study of the nonequilibrium dynamics induced by the optical excitation. 
The organic molecular solid  (TMTTF)$_2X$ ($X$: monovalent anion) is  a prototypical one-dimensional CO compound, in which the number of holes per the molecular orbital is 0.5.~\cite{ishiguro,jacobsen,balicas,dressel,monceau,naitoh}
The quantum fluctuation in a low dimensional lattice promotes a strong competition between the long-range CO insulator and the correlated metal, and governs the photoinduced transient dynamics between the two states. 
Quantum nature of lattice vibrations is also crucial on the energy transfer and the relaxation dynamics in the nonequilibrium transient states.~\cite{werner2,sayyad,murakami,golez,lee,dorfner,filippis}
From the theoretical viewpoint, several numerical algorithms have been applied for one-dimensional quantum systems.~\cite{maeshima05,maeshima06,maeshima07,iwano11,hassanieh}
Reliable calculations in one-dimensional systems may provide good references for the highly nonequilibrium transient dynamics in correlated electrons, in which the quantum effect, the many body effect and the transient characters are treated properly on an equal footing. 

In this paper, we address the photoinduced CO melting dynamics in a one-dimensional correlated electron system. 
In particular, we focus on roles of the quantum fluctuations of electrons and phonons on the transient dynamics. 
%To take into account of the quantum effects in the transient many body system, 
We analyze the one-dimensional spinless fermion model at half filling by using the infinite time-evolving block decimation (iTEBD) method. 
This method is well accepted as a reliable numerical algorithm for the one-dimensional interacting quantum model with high accuracy.
% not only for the ground state but also the transient states. 
In the model without the phonon degree of freedom, the CO melting is caused by soliton propagations,  corresponding to the domain-wall motions in the CO state. 
In the case of the finite fermion-phonon coupling, 
the transient dynamics depend qualitatively on the adiabatic parameter $\alpha \equiv \omega_0/t$, where $\omega_0$ and $t$ are the phonon frequency and the fermion hopping integral, respectively. 
In the case of large $\alpha$, the CO melting occurs in the same way to the case without the phonon degree of freedom and is induced by the soliton motions in which the velocity is renormalized by the polaronic effect. 
On the other hand, in the case of small $\alpha$, the CO melting occurs by the energy transfer to the phononic system at the first stage, and then by the soliton motion around the bottom of the soliton band. 
The present comprehensive analyses provide a standard reference for the photoinduced CO melting dynamics in low dimensional systems. 

In Sect.~II, the fermion-phonon coupled model is introduced, and the numerical method based on the iTEBD algorithm is briefly explained. 
In Sect.~IIIA, the numerical results for the ground state before photoexcitation are presented. 
The transient dynamics without the fermion-phonon interaction are introduced in Sect~IIIB, and the results with the phonon degree of freedom are presented in Sect.~IIIC. 
Section IV is devoted to discussion and summary. 
Detailed formulation for the classical soliton model is given in Appendix A.

\section{Model and Method}
The model in which we analyze the photoinduced CO dynamics is the interacting spinless-fermion model on a one-dimensional chain coupled with the local phonons. The Hamiltonian is defined as 
\begin{align}
\mathcal{H}=
&-\sum_{i} \left(t_{i, i+1} c_{i}^{\dagger}c_{i+1}+\text{H.c.} \right) \notag \\
&+V\sum_{i}\left(n_{i}-\frac{1}{2}\right)\left(n_{i+1}-\frac{1}{2}\right) \notag \\
&+\omega_{0}\sum_{i}a_{i}^{\dagger}a_{i}-g\sum_{i}(a_{i}+a_{i}^{\dagger})\left(n_{i}-\frac{1}{2}\right),
\label{eq:hamil}
\end{align}
where $c_{i} (c_{i}^{\dagger})$ is the annihilation (creation) operator of a spinless-fermion at site $i$,
$n_{i}(\equiv c_{i}^{\dagger}c_{i})$ is the fermion number operator, and 
$a_{i} (a_{i}^{\dagger})$ is the annihilation (creation) operator of  a local phonon at site $i$. 
The first term $({\cal H}_{t})$ and the second term $({\cal H}_{V})$ represent the fermion hopping and the Coulomb interaction between the fermions in the nearest neighboring sites, respectively. The third term $({\cal H}_{\rm ph})$ and the last term $({\cal H}_{\rm ep})$ represent the phonon energy and the Holstein-type fermion-phonon interaction with a coupling constant $g$, respectively. 
As already introduced in Sect.~I, $\alpha=\omega_0/t$ is the adiabatic parameter, 
and $\alpha \ll 1$ and $\alpha \gg 1$ are termed the adiabatic and antiadiabatic limits, respectively.    
We also introduce the dimension-less fermion-phonon coupling constant $\lambda \equiv g/\omega_0$. 

The light pulse applied along a chain direction is introduced as the Peierls phase in the hopping integral as 
\begin{align}
t_{i, i+1} \;\rightarrow\; t e^{{\rm i}A(\tau)} ,
\label{eq:pierls}
\end{align}
where the lattice constant, the light velocity and the Planck constant are taken to be one.  
The vector potential for the pump pulse at time $\tau$ is given as 
\begin{align}
A(\tau)=\frac{A_{p}}{\sqrt{2\pi}\tau_{p}}e^{\tau^{2}/(2\tau_{p}^{2})}\cos(\omega_{p}\tau),
\label{eq:vectorpotential}
\end{align}
where $A_{p}$, $\omega_{p}$, and $\tau_{p}$ are the amplitude, frequency and pulse width of the vector potential, respectively. The center of the vector potential is set to be $\tau=0$.

The wave functions for the fermions and phonons in the ground state and the time-evolved states are calculated using the iTEBD method.~\cite{vidal07,dorfner,takayoshi,ono1,ono2}
The wave function is represented as the matrix-product state form as 
\begin{align}
|\Psi \rangle=\sum_{\sigma_1, \sigma_2 ,\dots} \mathop{\textrm{Tr}} (B^{\sigma_1} B^{\sigma_2} \cdots ) 
|\sigma_1, \sigma_2, \dots \rangle , 
\end{align}
where $\sigma_i$ describes the local states for the fermion and phonon at site $i$, and $B^{\sigma_i}$ is a matrix with the dimension $\chi$.
The ground state wave function is obtained by the imaginary-time evolution as 
\begin{align}
|\Psi_{\rm GS} \rangle \propto \prod^{N}_{n=1} \exp(-{\cal H} \delta \tau) |\Psi_{0}\rangle ,
\label{eq:iteb}
\end{align}
where $|\Psi_0 \rangle$ is the initial wave function, and $\delta \tau$ is small difference of the imaginary time. 
%To calculate the exponential factor in Eq.~(\ref{eq:iteb}), we adopt the second-order Suzuki-Trotter decomposition given as $e^{A+B} \approx e^{A /2}e^{B}e^{A /2}$ where $A$ and $B$ are operators. 
In the calculation for the time-evolved states, $|\Psi (\tau)\rangle $, we use Eq.~(\ref{eq:iteb}), in which the imaginary time is replaced by the real time as $\delta \tau \rightarrow i \delta \tau$. 
To calculate the exponential factor in Eq.~(\ref{eq:iteb}), we adopt the second and fourth order Suzuki-Torotter decompositions for the imaginary-time and real-time equations, respectively.~\cite{hatano}
It is assumed that the wave functions are invariant under shifts of the two sites in most of the calculations, 
and have a periodicity under shifts of $64$ sites in the calculations for the local excitations introduced later. 
In order to monitor the CO state, we calculate the staggered CO parameter defined as 
\begin{align}
O=\frac{1}{N}\left| \sum_{i}(-1)^{i}\left( \langle n_{i}\rangle -\frac{1}{2}\right)\right|  , 
\label{eq:CO}
\end{align}
and the alternate ion displacement defined as 
\begin{align}
q=\frac{1}{N}\left| \sum_{i}(-1)^{i} 
\langle  a_i^\dagger+a_i   \rangle \right| . 
\label{eq:qlattice}
\end{align}
Here, $\langle \cdots \rangle$ represents the expectation value with respect to $|\Psi_{\rm GS }\rangle$ in the ground state, and that with respect to $|\Psi (\tau)\rangle$ for the time-evolved state. 
For each term of the Hamiltonian in Eq.~(\ref{eq:hamil}), 
we introduce the transient energy at $\tau$ measured from the initial energy defined as 
\begin{align}
\Delta E_X (\tau)=\langle \Psi(\tau)|{\cal H}_X|\Psi (\tau)\rangle-\langle \Psi_{\rm GS}|{\cal H}_X|\Psi_{\rm GS}\rangle , 
\end{align}
where $X$ identifies the term of the Hamiltonian. 

%----------------------------------------
\begin{figure}[t]
\includegraphics[width=0.7\columnwidth,clip]{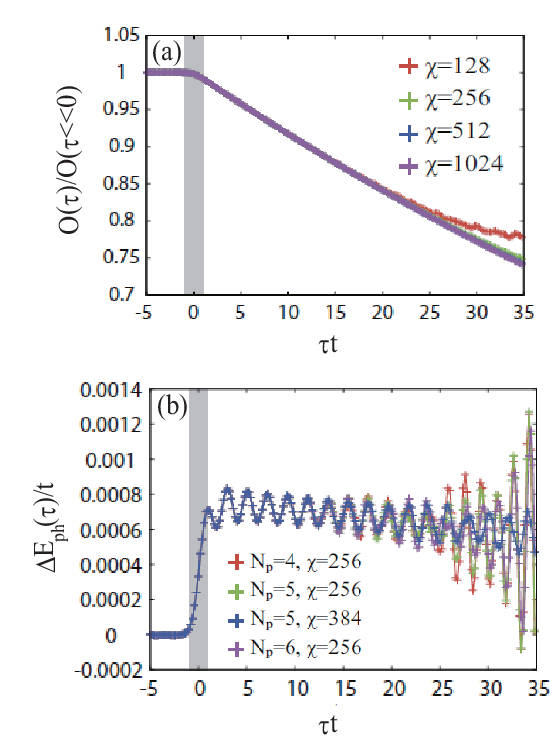}
\caption{
(Color online)
(a) Time dependences of the CO parameter for several values of $\chi$ at $V/t=6$ and $g=0$. Other parameters are chosen to be $\omega_p/t=4$, $A_p=0.1$ and $\tau_p=0.01$. 
(b) Time dependences of $\Delta E_{\rm ph}$ for several vales of $\chi$ and $N_{\rm ph}$.   
Other parameters are chosen to be $\alpha=3.2$, $\lambda=1.6$, $\omega_p/t=6$, $A_p=0.1$ and $\tau_p=0.01$. 
The shaded areas represent the time interval during which the pump pulse is introduced. 
}
\label{fig:accuracy}
\end{figure}
%----------------------------------------
Accuracies in the numerical calculations are checked by calculating the CO parameter and the ground state energies for various parameter sets. 
The present model at $g=0$, i.e. the interacting spinless fermion model without phonon, is equivalent to the $S=1/2$ XXZ model in a one-dimensional lattice, and $V=2t$ corresponds to the Heisenberg model.~\cite{mila,ejima}
The exact ground state is obtained by the Bethe ansatz:~\cite{samaj} 
a Luttinger-liquid state as the unique ground state in $V/t<2$ and the two-fold degenerated CO states in $V/t>2$. 
Differences of the calculated ground-state energy per site $(E_0)$ from the exact results $(E_{\rm exa})$, i.e.  $|(E_0-E_{\rm exa})/E_{\rm exa}|$, 
are of the order of $10^{-14}$ ($10^{-9} $) at $V/t=6$ ($V/t=3$) for $\chi=64$.  
Time dependences of the CO parameter at $g=0$ and $V/t=6$ are shown in Fig.~\ref{fig:accuracy}(a) for various values of $\chi$. It is confirmed that the data for $\chi \ge 256$ converge up to $\tau t \simeq 30$. 
In the calculations for $g \ne 0$, the maximum phonon number per site $(N_{\rm ph})$ is truncated. 
In Fig.~\ref{fig:accuracy}(b), $\Delta E_{\rm ph}$ are shown for several values of $N_{\rm ph}$ and $\chi$. 
The calculated results are almost converged up to around $\tau t=25$ at $N_p=5$ and $\chi=256$. 
By taking into account of the above results, in most of the numerical calculations for the transient states,  
we adopt $\chi=512$ for $g/\omega_0 < 0.5$, $\chi=384$ for $0.5 \leq g/\omega_0 < 1$, and 
$\chi=256$ for $1 \leq g/\omega_0 $, and $N_{\rm ph}=3-8$.

The excitation spectra are calculated based on the matrix-product state formalism, in which finite size clusters of $L_c$ sites with the open boundary condition are adopted. 
Here, the spectral functions are expanded using the Chebyshev polynomial.~\cite{holzner} 
We define the excitation spectra induced by the operator ${\cal A}$ as 
\begin{align}
\chi_{{\cal AA}}(\omega)= \langle \Psi_{\rm GS} \left | {\cal A}^\dagger \delta(\omega-{\cal H}+E_0) {\cal A} \right | \Psi_{\rm GS} \rangle . 
\end{align} 
The spectral function is expanded as 
\begin{align}
\chi_{{\cal AA}}(\omega)=\frac{1}{\pi \sqrt{1-\omega^2}}
\left [  g_0 \mu_0+2 \sum^{N_{\rm ch}-1}_{n=1} g_n \mu_n T_n (\omega)        \right ] , 
\end{align}
where $T_n(\omega)$ is the $n$-th Chebyshev polynomial, 
$\mu_n = \langle \Psi_{\rm GS}|  {\cal A}^\dagger  T_n({\cal H}) {\cal A} |\Psi_{\rm GS}\rangle$ is the Chebyshev moment, and $g_n$ is the correction factor defined by 
\begin{align}
g_n=\frac{(N_{\rm ch}-n+1)\cos\frac{\pi n}{N_{{\rm ch}}+1} + \sin \frac{\pi n}{N_{\rm ch}+1} \cot \frac{\pi}{N_{\rm ch}+1}}
{N_{\rm ch}+1} . 
\end{align}
Based on the recursion formula for the Chebyshev vectors, $|t_{n+1}\rangle=2{\cal H}|t_n\rangle-|t_{n-1}\rangle$ with $|t_0\rangle={\cal A}|\Psi_{\rm GS} \rangle$ and $|t_1\rangle={\cal H}|t_0 \rangle$, these are calculated in the matrix-product state formalism.

\section{Results}

\subsection{Ground state}
%----------------------------------------
\begin{figure}[t]
\includegraphics[width=\columnwidth,clip]{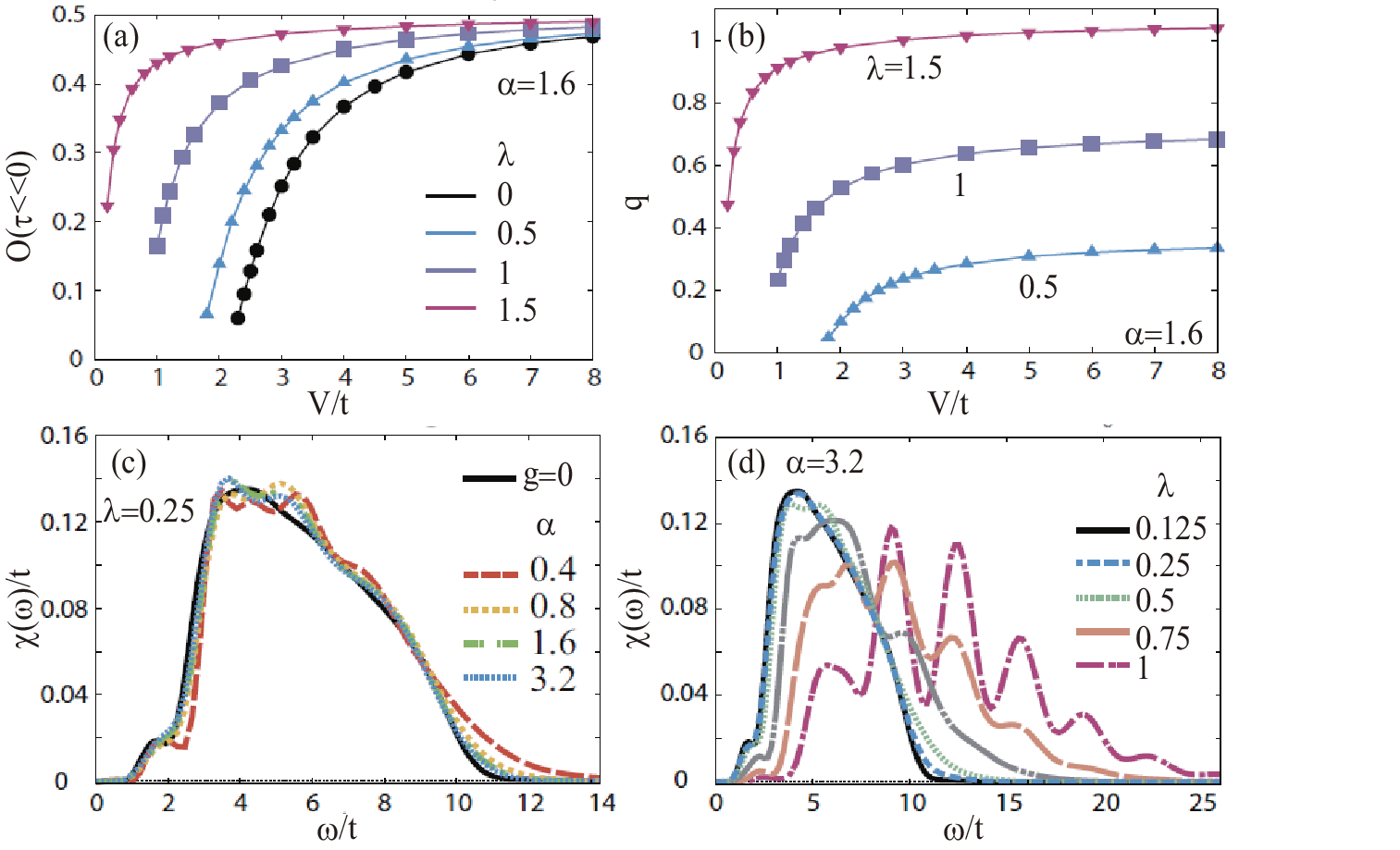}
\caption{
(Color online)
(a) The CO parameter and (b) the alternate ion displacement for several values of $V/t$ and $\lambda$ in the ground state. 
Other parameter value is chosen to be $\alpha=1.6$. 
(c) The optical absorption spectra at $\lambda=0.25$ for several values of $\alpha$, and (d) the spectra at $\alpha=3.2$ for several values of $\lambda$. 
The results at $g=0$ are also shown in (c) for comparison. 
Other parameters are chosen to be $V/t=6$ and $L_c=64$ in (c) and (d). 
}
\label{fig:gs}
\end{figure}
%-----------------------------------------
%
Before showing the photoinduced dynamics in the CO systems, we introduce the ground state properties as the  initial state. 
In Figs.~\ref{fig:gs}(a) and (b),  the CO parameter defined in Eq.~(\ref{eq:CO}) and the lattice distortions in Eq.~(\ref{eq:qlattice}) for several values of $g$ are presented, respectively. 
Without the fermion-phonon coupling ($g=0$), the CO parameter tends to vanish at $V/t=2$ with decreasing $V/t$ as predicted by the exact solution.  
With increasing $g$, the CO phase is strengthened. 
The alternate ion distortion occurs in the CO phase, and the saturated values of $q$ increase with $g$. 

The optical absorption spectra are shown in Fig.~\ref{fig:gs}(c) for several values of $\alpha$ as well as $g=0$, and in Fig.~\ref{fig:gs}(d) for several values of $\lambda$. 
We define the optical absorption spectra as 
\begin{align}
\chi_{jj}(\omega)=\langle \Psi_{\rm GS} \left | j \delta (\omega-{\cal H}+E_0) j \right | \Psi_{\rm GS} \rangle , 
\end{align}
where 
$j=\sum_i ({\rm i} t c_i^\dagger c_{i+1}+{\rm H.c.} )$ is the current operator along the chain. 
The elementary excitations in the ordered phase of the one-dimensional XXZ model is known as the domain wall motions. The optical absorption spectra, in which the net momentum transfer is zero, are ascribed to the kink and antikink pair with the opposite momenta.~\cite{ishimura,kanamori} 
The energy for a non-interacting pair of the kink with momentum $k$ and the antikink with momentum $k'$ is given by 
 \begin{align}
E(k,k')=\varepsilon_{k}+\varepsilon_{k'} , 
\label{eq:solitonenergy}
 \end{align}
with $\varepsilon_k=-2t \cos(2k)+V/2$. 
This formula indicates that the optical absorption spectra distribute in $V-4t \le \omega \le V+4t$, which is 
consistent with the main part of the optical spectra shown in Fig.~\ref{fig:gs}(c) where $V/t=6$. 
We have confirmed that a weak hump structure below $V-4t$, such as a small peak at around $\omega/t=1.9$, reduces with increasing the cluster size, and this is an artifact due to the size effect. 
On the other hand, the main structure does not depend sensitively on the cluster size. 
The spectra in the antiadiabatic case ($\alpha=3.2$) are similar to the spectra without the fermion-phonon interaction, and some fine structures appear in the adiabatic case ($\alpha=0.4$). 
With increasing $\lambda$, multiple peak structures appear above the main absorption edge at around $V+n \omega_0$ with an integer number $n$ as shown in Fig.~\ref{fig:gs}(d). 
In the calculations of the photoinduced dynamics shown in the following section, the pump-photon frequency is chosen around the energy in which $\chi_{jj}(\omega)$ is finite, implying that the pump photon yields  resonant excitations.

\subsection{Transient state without the fermion-phonon coupling} 
\label{sec:without_ep}
%--------------------------------------------------------------------------------------------------
\begin{figure}[t]
\includegraphics[width=0.7\columnwidth,clip]{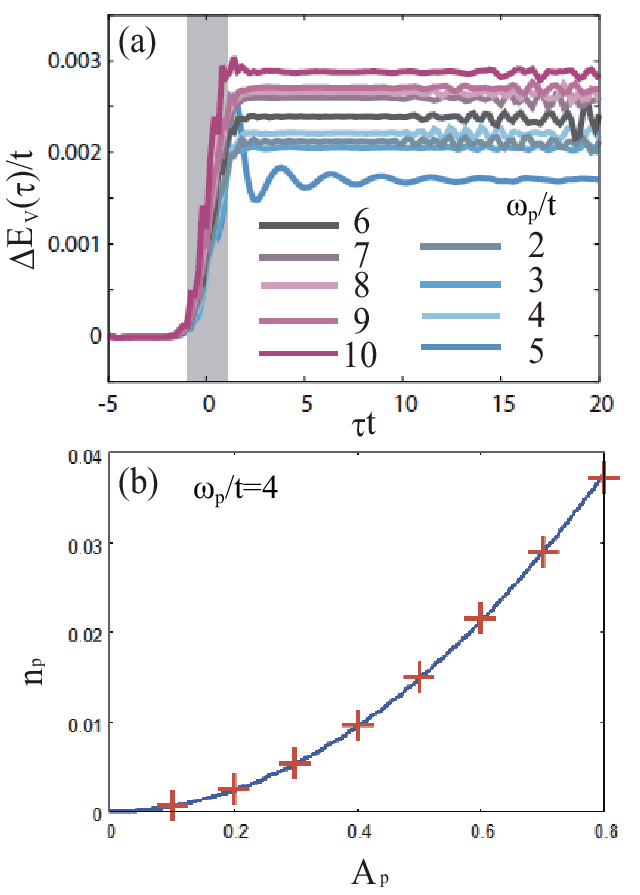}
\caption{
(Color online)
(a) Time dependence of $\Delta E_V$ for several values of $\omega_p$. 
The shaded areas represent the time interval during which the pump pulse is introduced. 
(b) The effective absorbed photon number per site at $\omega_p/t=4$. 
Other parameter values are chosen to be $V/t=6$ and $g=0$. 
}
\label{fig:energy_g0}
\end{figure}
%--------------------------------------------------------------------------------------------------
In this subsection, the photoinduced CO melting in the case without the fermion-phonon interaction are introduced. 
In Fig.~\ref{fig:energy_g0}(a), the time profiles of $\Delta E_V(\tau)$ after the excitation are shown for several values of $\omega_p$.
We introduce the effective absorbed photon number per site defined as
$ n_p=N_p (\tau \gg \tau_p)$
with  
\begin{align}
N_p(\tau)=\frac{\Delta E(\tau )}{N\omega_p} , 
\label{eq:np}
\end{align}
where 
$\Delta E (\tau)=\langle \Psi(\tau)|{\cal H}|\Psi (\tau)\rangle-E_0$ is a change in the total energy. 
In several cases of $\omega_p$ adopted in Fig~\ref{fig:energy_g0}(a), $n_p$ are unified to be approximately 0.0005 by adjusting $A_p$. 
 After introducing the pump pulse, $\Delta E_V(\tau)$ rapidly increases and settles at almost time-independent values; the photoexcited state is stable after turning off the pump pulse. 
In the case of small $\omega_p$, a oscillatory behavior appears and is damped  at around $\tau=10/t$. 
As shown in Fig.~\ref{fig:energy_g0}(b), $n_p$ increases quadratically as a function of $A_p$. 
This is interpreted that the number of the domain walls in the CO states, i.e. the kinks and antikinks, increases with $A_p$. 
%With increasing $\omega_p$, $\Delta E_V$ monotonically increases, implying creation of the the kinks (antikinks) in the CO state. 
%

%--------------------------------------------------------------------------------------------------
\begin{figure}[t]
\includegraphics[width=0.7\columnwidth,clip]{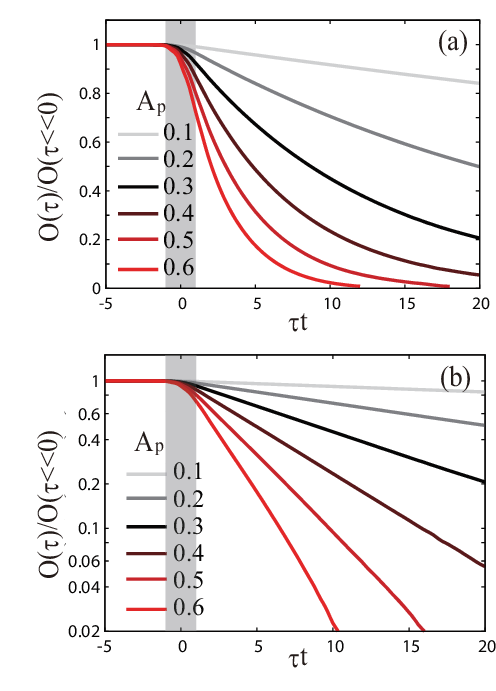}
\caption{
(Color online)
(a) Time dependences of the normalized $O(\tau)$, and (b) the semi-logarithmic plot of $O(\tau)$ for several values of $A_p$. 
The shaded areas represent the time interval during which the pump pulse is introduced. 
Other parameter values are chosen to be $V/t=6$, $\omega_p/t=4$ and $g=0$. 
}
\label{fig:order_g0}
\end{figure}
\begin{figure}[t]
\includegraphics[width=0.7\columnwidth,clip]{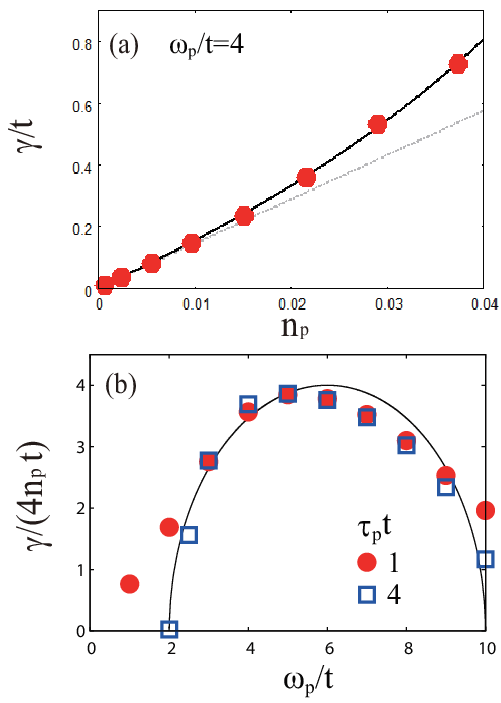}
\caption{
(Color online)
(a) $n_{p}$ dependence of $\gamma$ at $\omega_p/t=4$. 
Solid and broken lines represent Eq.~(\ref{eq:classical}) and 
$\gamma=4 v n_p$, respectively.
(b) $\omega_{p}$ dependence of $\gamma/(4n_{p})$ at $A_p=0.1$. 
A solid line represents Eq.~(\ref{eq:vel_ene}).
Other parameter is chosen to be $V/t=6$. 
}
\label{fig:gamma_g0}
\end{figure}

%--------------------------------------------------------------------------------------------------
%
The time profiles of the CO parameter are shown in Fig.~\ref{fig:order_g0}(a).
Reduction in $O(\tau)$ occurs monotonically for all values of $A_p$. 
In contrast to the time dependence of $\Delta E_V(\tau)$ shown in Fig.~\ref{fig:energy_g0}(a), the CO melting occurs in a long time scale after turning off the pump pulse. 
In the case of large $A_p$, $O(\tau)$ reaches zero, implying almost complete melting of the initial CO. 
As shown in the semi-logarithmic plot in Fig.~\ref{fig:order_g0}(b), 
the time dependences of $O(\tau)$ follows an exponential function as 
\begin{align}
O(\tau) =O_0 \exp(-\gamma\tau), 
\label{eq:exp}
\end{align}
where $O_0$ is the CO parameter in the initial state, and $\gamma$ is the damping factor.
We have confirmed that the exponential dependence of $O(\tau)$ is commonly observed in wide parameter ranges of $\omega_p$ and $V$. 
The damping factor is calculated for several values of $A_p$ at fixed $\omega_p(=4t)$, 
and is plotted as a function of $n_p$ in Fig.~\ref{fig:gamma_g0}(a). 
It increases linearly for small $n_p$, and deviates upward with increasing $n_p$.  
As shown in Fig.~\ref{fig:gamma_g0}(b), the $\omega_p$ dependence of $\gamma$ shows a dome shape. 
%This behavior does not depend on the pulse width $\tau_p$. 
These two behaviors of $\gamma$ are well explained by the results of the soliton picture plotted by bold lines in Figs.~\ref{fig:gamma_g0}(a) and (b), as explained later. 

%-----------------------------------------------------------------------------
%-----------------------------------------------------------------------------

\begin{figure}[t]
\includegraphics[width=0.7\columnwidth,clip]{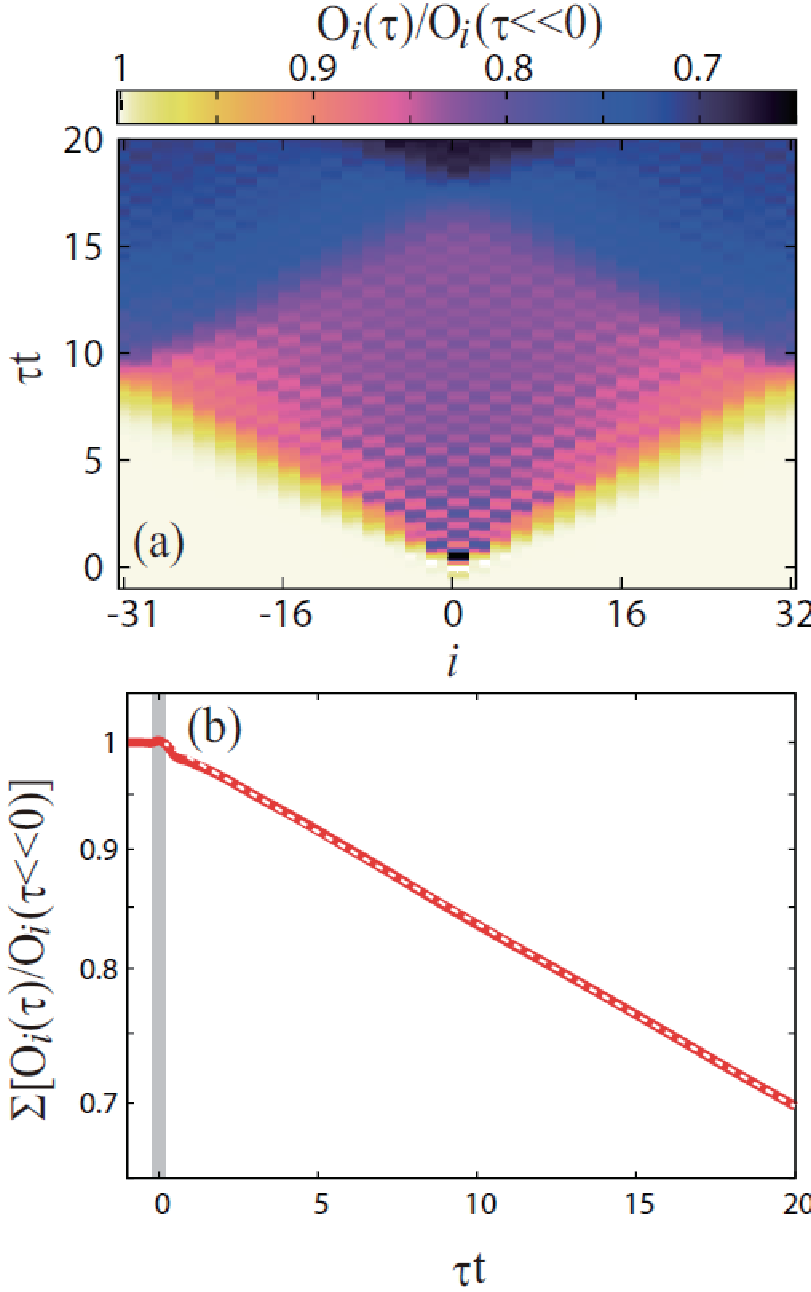}
\caption{
(Color online)
Numerical results induced by the local excitation. 
(a) An intensity plot of $O_{i}(\tau)/O_i(\tau \ll 0)$ in the space-time plane.
(b) A semi-logarithmic plot of the time profile of $\sum_{i} [O_i(\tau)/O_i(\tau \ll 0)]$. 
The shaded areas represent the time interval during which the pump pulse is introduced. 
Other parameter values are chosen to be $V/t=6$, $g/t=0$, $\omega_p/t=6$, $A_p=0.8$, $\tau_pt=0.2$ and $L=64$. 
}
\label{fig:local_g0}
\end{figure}

The CO melting dynamics are analyzed by the following artificial electronic excitation. 
The vector potential for the pump pulse is applied at a specific site  
instead of the uniform vector potential introduced in Eq.~(\ref{eq:pierls}). 
This is termed ``the local excitation" from now on and is defined as 
\begin{align}
t_{i, i+1} \rightarrow
\begin{cases}
t e^{{\rm i}A(\tau)} \ \   & (i=\cdots -L, 0, L, \cdots) \\
t \ \   & \text{other sites},
\end{cases}
\label{eq:local}
\end{align}
where $L$ is a periodicity of the sites in which the excitations are induced, and the time dependence of $A(\tau)$ is given in Eq.~(\ref{eq:vectorpotential}). 
Responses to the local excitation are monitored by the CO parameter defined at each site termed ``the local CO parameter" defined as
\begin{align}
O_{i}(\tau)=(-1)^{i}\left (\Braket{n_{i}}-\frac{1}{2}\right ) .
\end{align}
An intensity plot of $O_i(\tau)$ induced by the local excitations in a space-time plane is plotted in Fig.~\ref{fig:local_g0}(a). 
A periodicity of the local excitations is chosen to be $L=64$.
After introducing the pump pulse, 
a CO domain characterized by the reduced $O_i(\tau)$ is generated, and the boundaries between the domains   propagate almost linearly with time. 
At $\tau\approx 9/t$, 
the domain wall collides at $i=\pm 32$ with other domain walls which are generated at $i=\pm L$. 
After the collision, $O_i$ reduces furthermore. 
Next collisions occur at  approximately $\tau= 18/t$ and $i=0$.
A velocity of the domain wall is about 4 sites per the unit time of $1/t$. 
After the collisions, both the widths and velocities of the domain walls are almost unchanged. 
From these characteristics, the domain wall dynamics in the photoinduced CO state are identified as the soliton motions. 
As plotted in Fig.~\ref{fig:local_g0}(b), the CO parameter $O(\tau)$ calculated from the data of $O_i (\tau)$
shows the exponential decay in the same way to the results in Fig.~\ref{fig:order_g0}. 
We conclude that the soliton motions are responsible for the photoinduced CO melting. 
%The $n_{p}$ dependence of the $O_{i}(\tau=10)$ at $i=0$ is shown in Fig.~\ref{fig:local_g0}.
%The $O_{i}$ after the local excitation decreases linearly with $n_{p}$.

%--------------------------------------------------------
%--------------------------------------------------------
%
Next, we derive the damping factor based on the soliton picture. 
We set up the model in which the photogeneraged solitons are treated as classical particles. 
A photoinduced fermionic excitation is simulated by a stochastic generation of solitons. 
We assume that the soliton velocity is constant $v$, and the collisions between the solitons are elastic. 
Details are presented in Appendix A.
We obtain an expression of $\gamma$ as 
\begin{align}
\gamma=-\frac{2v\ln(1-2Ln_{p})}{L} , 
\label{eq:classical}
\end{align}
which is approximately $\gamma=4vn_p$ in the limit of $n_p \ll 1$. 
The soliton velocity in Eq.~(\ref{eq:classical}) is evaluated from the effective Hamiltonian derived from Eq.~(\ref{eq:hamil}) in the limit of $V \gg t$. 
The number of the soliton-antisoliton pair is assumed to be one. 
The effective Hamiltonian is given as~\cite{kanamori}
\begin{align}
{\cal H}_{\rm eff}=-t\sum_{i={\rm odd}} \left (b_i^\dagger b_{i+2}+d_i^\dagger d_{i+2} +{\rm H.c.}    \right ) 
+\frac{V}{2} , 
\label{eq:solitonmodel}
\end{align}
where $b_i^\dagger$ ($b_i$) and $d_i^\dagger$ ($d_i$) are the creation (annihilation) operators of the hard-core bosons for the soliton and antisoliton, respectively. 
A summation with respect to $i$ takes for odd integers. 
The interaction between the soliton and antisoliton given by the second oder of $t$ is not included. 
The eigenstates are classified by the momenta of the soliton $k$ and antisoliton $k'$, 
and the eigenenergy coincides to Eq.~(\ref{eq:solitonenergy}). 
The group velocity of soliton is obtained as 
\begin{align}
v_{k}=|4t\sin 2k|. 
\label{eq:velocity}
\end{align}
A factor 2 in the sine function is attributed to the fact that the fermion hopping between the nearest neighbor  site corresponds to the soliton hopping between the next-nearest neighbor sites. 
In the case where the photon momenta is zero and the resonant excitation is induced, i.e. $\omega_p=E(k, -k)$, the velocity is represented as a function of $\omega_p$ as  
\begin{align}
v_{\omega_p}=\sqrt{(4t)^{2}-(\omega_p-V)^{2}} . 
\label{eq:vel_ene}
\end{align}%
The maximum group velocity is given as $4t$, being consistent with the results in Fig.~\ref{fig:local_g0}(a). 

The analytical expression of the damping factor in Eq.~(\ref{eq:classical}) with Eq.~(\ref{eq:vel_ene}) is compared with the numerical data. 
Here, $L$ in Eq.~(\ref{eq:classical}) is treated as a fitting parameter and is determined to be approximately $L=6.5$. 
It is shown that the $n_p$ dependence of $\gamma$ plotted in Fig.~\ref{fig:gamma_g0}(a) well reproduces the calculated data. 
As shown in Fig.~\ref{fig:gamma_g0}(b),  $\gamma$ as a function of $\omega_p$ shows a dome shape structure which takes its maximum at approximately $\omega_p=V$. 
This behavior is confirmed by the numerical data for several values of $V$.  
The analytical formulation little deviates from  the numerical data in small $V$, in which 
the higher-order quantum fluctuation neglected in Eq.~(\ref{eq:classical}) might be important. 

%-----------------------------------------------------------------------------------------
\begin{figure}[t]
\includegraphics[width=\columnwidth,clip]{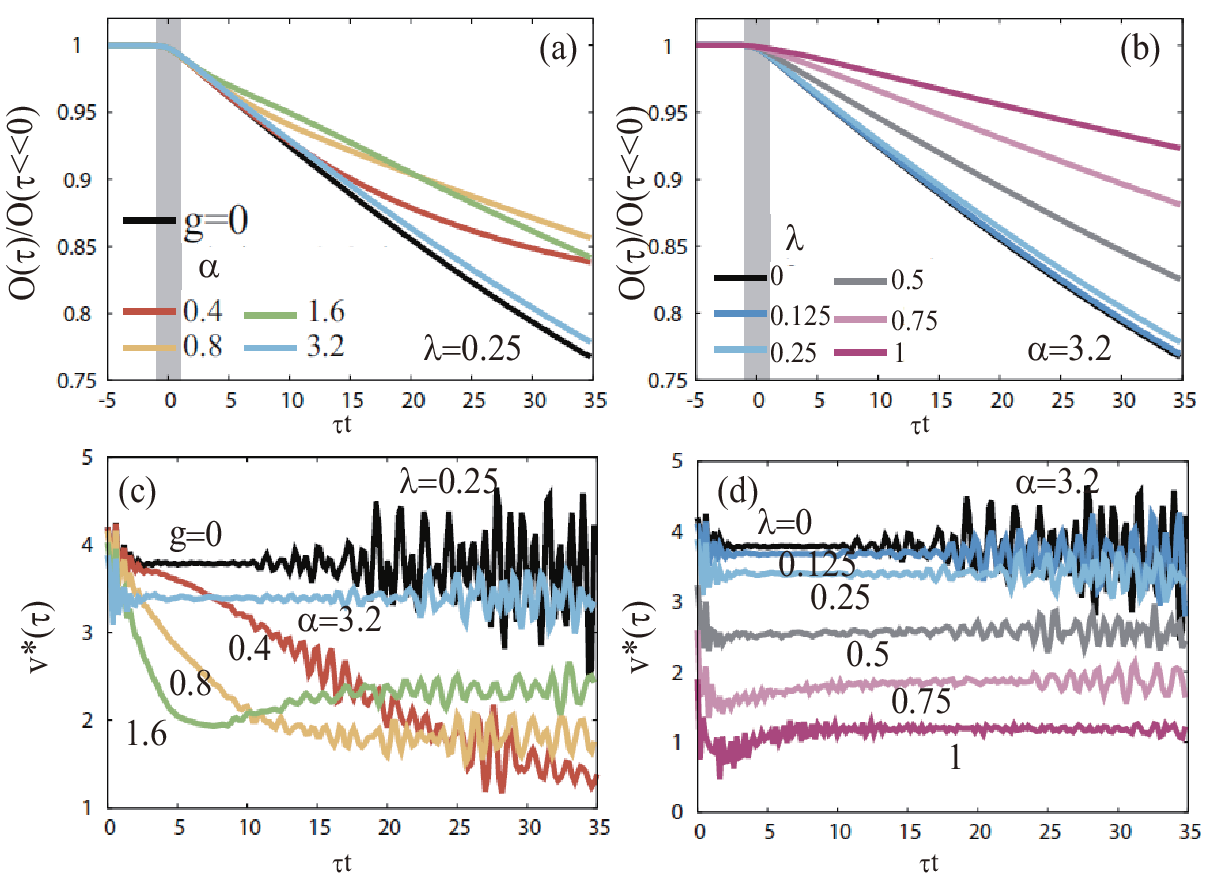}
\caption{
(Color online)
Time profiles of $O(\tau)$ and $v^\ast(\tau)$ for (a, c) several values of $\alpha$ at $\lambda=0.25$,  
and (b, d) several values of $\lambda$ at $\alpha=3.2$. 
Results at $\lambda=0$ are also plotted for a comparison in (a, c). 
%Time profiles of $O(\tau)$ and $v^\ast(\tau)$ for several values of $\lambda$ at $\alpha=3.2$. 
The shaded areas represent the time interval during which the pump pulse is introduced. 
Other parameter values are chosen to be $V/t=6, \omega_p/t=6$, and $\tau_{p}t=1$.  
%\red{(thin lines'ð}'ɒljÁ)}
}
\label{fig:order_g}
\end{figure}
%-----------------------------------------------------------------------------------------

\subsection{Effect of the fermion-phonon coupling}
In this subsection, the numerical results in the case of $g \neq 0$ are presented. 
Figure~\ref{fig:order_g}(a) shows the time profiles of $O(\tau)$ 
for several values of $\alpha$ at $\lambda=0.25$.  
For comparison, we also present the results at $g=0$, in which $O(\tau)$ decreases exponentially as a function of time as explained previously. 
With introducing the fermion-phonon coupling, the reduction of $O(\tau)$ is suppressed. 
The exponential decay of $O(\tau)$ remains in the case of large phonon frequency  $(\alpha=3.2)$, but the time profiles of $O(\tau)$ deviates from the exponential decay with decreasing $\alpha$. 
To clarify the change in the CO melting dynamics by the fermion-phonon coupling, 
the concept of the soliton velocity introduced previously is extended to the transient state.  
From Eqs.~(\ref{eq:exp}) and (\ref{eq:classical}) under the conditions of $\gamma \tau \ll 1$ and $n_p \ll 1$,  we have $O(\tau) \sim e^{-\gamma \tau} \sim 1-4vn_p \tau$. 
Then, the transient soliton velocity at time $\tau$ is introduced as 
\begin{align}
v^{\ast}(\tau)=-\frac{1}{4N_{p}(\tau)}\frac{dO(\tau)}{d\tau},
\end{align}
where $N_p(\tau)$ is defined in Eq.~(\ref{eq:np}). 
The calculated results of $v^\ast(\tau)$ are shown in Fig.~\ref{fig:order_g}(c). 
%, in which $g=0$ and $g/\omega_0=0.25$ with several $\omega_0$ are chosen. 
In the case of the large $\alpha$ ($\alpha=3.2$), $v^\ast (\tau)$ is almost independent of time after turning off the pump pulse in the same way of $g=0$. 
The oscillatory behaviors seen in $\tau \geq 10$ are suppressed with increasing $\chi$, and are attributed to  the numerical artifact. 
On the other hand,  for the small $\alpha$ $(\alpha=0.4, {\rm and \ } 0.8)$, $v^\ast(\tau)$ changes remarkably in a short time scale and tends to be saturated in a long time scale. 
%
%The thin lines in Fig.~\ref{fig:order_g}(c) represent the soliton velocity obtained by fitting the numerical data of $O(\tau)$ in the time domain where using Eqs.~(\ref{eq:exp}) and (\ref{eq:classical}), implying that the time profiles of $O(\tau)$ follow the exponential dependence in this time domain. 
This characteristic time scale in the adiabatic case ($\alpha=0.4$ and 0.8) will be discussed later.  
The time profiles of $O(\tau)$ for several values of $\lambda$ are presented in Figs.~\ref{fig:order_g}(b) and (d), where $\alpha$ is fixed to be 3.2, i.e. the antiadiabatic condition. 
The exponential time dependence of $O(\tau)$ and the almost constant $v^\ast(\tau)$ are observed for $0 \le \lambda \le 3.2$, and $v^\ast(\tau)$ monotonically decreases with increasing $\lambda$. 
Detailed CO melting dynamics in the cases of the adiabatic and antiadiabatic conditions are explained in the following two subsections.  

\subsubsection{Antiadiabatic case ($\alpha \gsim 1$)}

%-------------------------------------------------------------------------------
\begin{figure}[t]
\includegraphics[width=\columnwidth,clip]{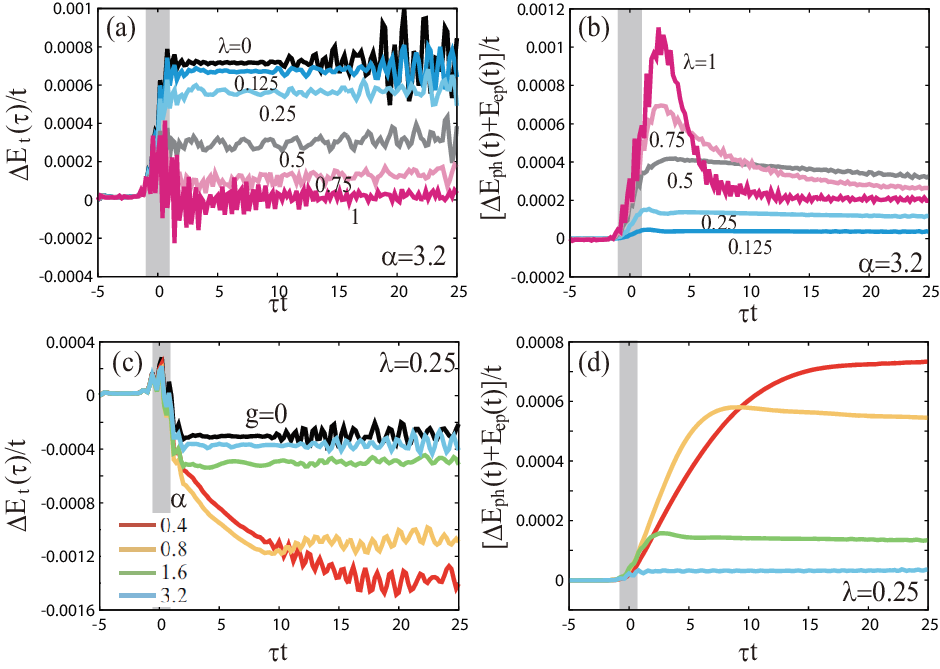}
\caption{
(Color online)
Time profiles of $\Delta E_t(\tau)$ and $\Delta E_{\rm ph}(\tau)+\Delta E_{\rm ep}(\tau)$ for (a, b) several values of $\lambda$ at $\alpha=3.2$, and (c, d) several values of $\alpha$ at $\lambda=0.25$. 
Results at $\lambda=0$ are also plotted for a comparison in (a, c). 
The shaded areas represent the time interval at which the pump pulse is introduced.
Other parameter values are chosen to be $V/t=6$ and $\omega_p/t=6$.
}
\label{fig:energy_g}
\end{figure}
%-------------------------------------------------------------------------------

\begin{figure}[t]
\includegraphics[width=\columnwidth,clip]{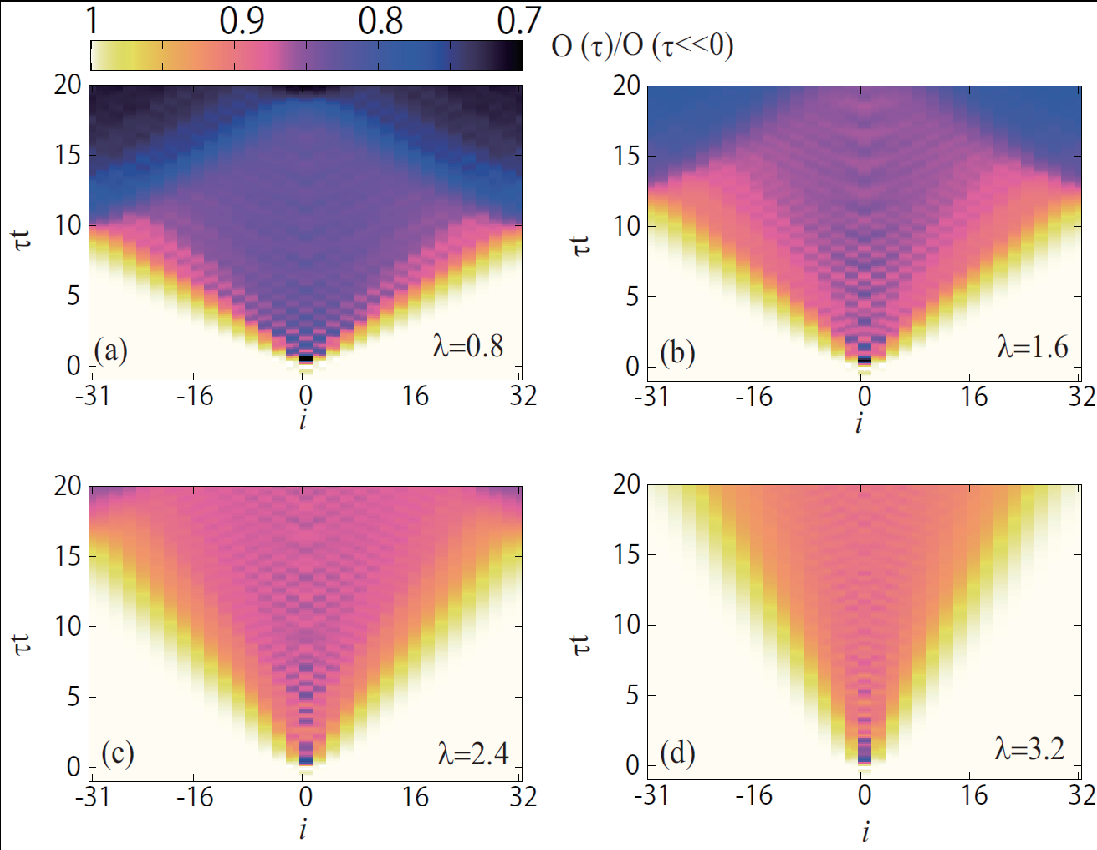}
\caption{
(Color online)
Intensity plots of $O_{i}(\tau)$ in the space-time plane for several $\lambda$ at $\alpha=3.2$ 
induced by the local excitations. 
Other parameter values are chosen to be $V/t=6, \omega_{p}/t=6$, and $\tau_{p}t=0.2$.
}
\label{fig:local_g}
\end{figure}

%We have shown that the CO melting dynamics are crucially influenced by the phonon frequency. 
First, we focus on the dynamics in the antiadiabatic case at $\alpha=3.2(>1)$, where the CO parameter decays exponentially. 
The time profiles of $\Delta E_t(\tau)$ and $\Delta E_{\rm ph}(\tau)+\Delta E_{\rm ep}(\tau)$ are shown for several values of $\lambda$ in Figs.~\ref{fig:energy_g}(a) and (b), respectively. 
Except for small dip and hump structure just after turning off the pump pulse at $\lambda=2.4 \ {\rm and } \ 3.2$, 
both of the energies do not show remarkable time dependences.  
With increase $\lambda$, the saturated value of $\Delta E_t(\tau)$ decreases monotonically. 
%In the cases of $g=2.4$ and $3.2$, dip structures in $\Delta E_{ph}+\Delta E_{ep}$ appear around $\tau=3/t$, and are smoothly saturated at constant values. 
%
%Roles of the electron-phonon coupling on the photoinduced CO melting in the case of $\omega_0$, 
The intensity plots in the space-time plane for $O_i(\tau)$ induced by the local excitation are presented in Fig.~\ref{fig:local_g}. 
%, where the local order parameters $O_{i}(\tau)$ for several values of $g$ at $\omega_{0}/t=3.2$ are plotted. 
For all cases of $\lambda$, the soliton motions are stable and their velocities are almost unchanged. 
With increasing $\lambda$, the velocity decreases, being consistent with the results in Fig.~\ref{fig:order_g}(d). 
% and is identified as a reason of the CO melting suppression. 

%-------------------------------------------------------------------------------
%-------------------------------------------------------------------------------
\begin{figure}[t]
\includegraphics[width=0.7\columnwidth,clip]{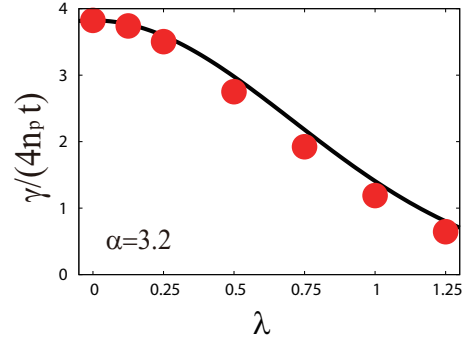}
\caption{
(Color online)
The normalized damping factor $\gamma/(4n_{p})$ as a function of $\lambda$ at $\alpha=3.2$. 
A solid line represents Eq.~(\ref{eq:polaron}).
Other parameter values are chosen to be $V/t=6, \omega_{p}/t=5$, and $\tau_{p}t=1$.
}
\label{fig:gamma_g2}
\end{figure}
%-------------------------------------------------------------------------------

From the numerical data shown in Fig.~\ref{fig:order_g}, the normalized damping rate $\gamma/(4n_pt)$, corresponding to the soliton velocity, is plotted as a function of $\lambda$ in Fig.~\ref{fig:gamma_g2}.
%Here, we chose $\omega_p=3.2$ where the $v_s$ is maximum as a function of $\omega_p$
The velocity decreases monotonically with increasing $\lambda$. 
The bold line in the figure represents the equation given by 
\begin{align}
\gamma/(4n_p t) \propto e^{-g^2/\omega_0^2} , 
\label{eq:polaron}
\end{align}
which is expected from the renormalized carrier hopping in the polaronic state based on the Lang-Firsov theory.~\cite{lang} 
The analytical formula fits well the numerical data even in the region of small $\lambda$, implying that  the coherent soliton motion associated with the ion displacement occurs from weak to strong coupling regimes. 
This is termed that polaronic-soliton picture from now on, and is consistent with the fact that $\Delta E_{\rm ph}(\tau)$ does not show remarkable change after turning off the pump pulse (not shown). 
We conclude that, in the antiadiabatic case, the CO melting dynamics is caused by the soliton motion renormalized by the polaron formation. 

%\begin{figure}[t]
%\includegraphics[width=0.7\columnwidth,clip]{fig/energy_f.eps}
%\caption{
%(Color online)
%Time dependence of (a) absorbed photon densities, (b) kinetic energies, (c) Coulomb energies,(d) phonon energies, (e) electron--phonon energies, and (f) sum of phonon- and electron--phonon- energies.The shaded areas represent the time interval at which the pump pulse is introduced.The parameter values are the same with those in Fig.~\ref{fig:order_g}(a).
%}
%\label{fig:energy_f}
%\end{figure}
%

\subsubsection{Adiabatic case ($\alpha \lsim 1$)}

Next, we focus on the CO melting dynamics in the adiabatic case in $\alpha<1$, where the time dependence of $O(\tau)$ deviates from the exponential decay as shown in Fig.~\ref{fig:order_g}(a). 
The time profiles of $\Delta E_t(\tau)$ and $\Delta E_{\rm ph}(\tau)+\Delta E_{\rm ep}(\tau)$ are shown for several values of $\alpha$ in Figs.~\ref{fig:energy_g}(c) and (d), respectively. 
It is confirmed that $\Delta E_V(\tau)$ suddenly increases by pulse irradiation, and does not show remarkable 
time dependence after turning off the pump pulse (not shown). 
The time profiles of $\Delta E_t(\tau)$ are clearly different from those in the antiadiabatic case presented in
Fig.~\ref{fig:energy_g}(a); 
$\Delta E_t(\tau)$ gradually decreases after turning off the pump pulse and are saturated at approximately  $\tau t=10$ and 15 for $\alpha=0.8$ and 0.4, respectively. 
Counter behaviors are observed in $\Delta E_{\rm ph}(\tau)+\Delta E_{\rm ep}(\tau)$, which increases  gradually with time. 
We also confirm that $\Delta E_{\rm ep}(\tau)$ is one order smaller than $\Delta E_{\rm ph}(\tau)$. 
The above results are interpreted that the excess energy by the pump excitation is transfered from the fermionic to phononic systems, and the charge distributions are compatible to the ion displacements. 

%---------------------------------------------------------------------------------------
\begin{figure}[t]
\includegraphics[width=0.8\columnwidth,clip]{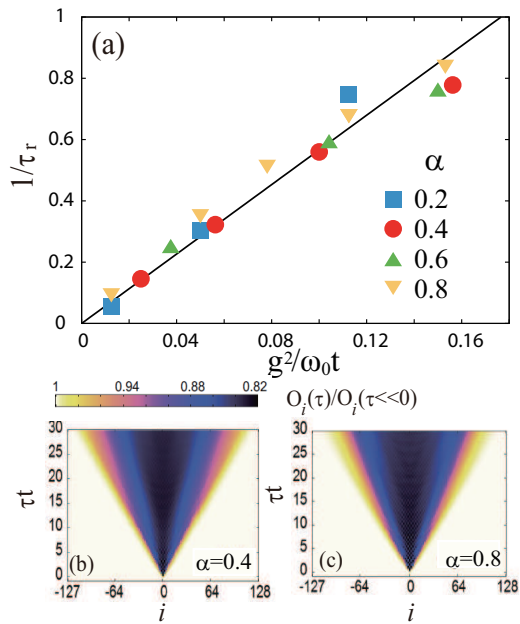}
\caption{
(Color online)
(a) The inverse of the energy relaxation time as a functions of $g^2/(\omega_0 t)$. 
Data are obtained by fitting $\Delta E_{\rm ph}(\tau)+\Delta E_{\rm ep}(\tau)$ for several values of $\alpha$ using the equation $C (e^{-\tau /\tau_r}-1)$. 
 Other parameters are chosen to be $V/t=6$, $\omega_p/t=4$ and $\tau_p t=1$. 
A solid straight line is for guide of eye. 
(b, c) Intensity plots of $O_i(\tau)$ in the space-time plane induced by the local excitation at $\alpha=0.4$ and $0.8$ with $\lambda=0.25$. 
Other parameters are chosen to be $V/t=6$, $\omega_p/t=6$ and $\tau_p t=0.2$. 
}
\label{fig:relax}
\end{figure}
%---------------------------------------------------------------------------------------
%
In order to analyze the energy transfer, the time profiles of $\Delta E_{\rm ph}(\tau)+\Delta E_{\rm ep}(\tau)$ are fitted by a function $C (e^{-\tau /\tau_r}-1)$ where $\tau_r$ is the energy relaxation time, and $C$ is a numerical constant. 
In Fig.~\ref{fig:relax}, the energy relaxation times calculated for several values of $\alpha(<1)$ and $\lambda$ are plotted as a functions of $g^2/(\omega_0 t)$. 
Almost linear dependence is observed until $g^2/(\omega_0 t) \sim 0.16$,
implying the energy relaxation in accordance with the Fermi golden rule through the fermion-phonon coupling.~\cite{werner2,sayyad,murakami,golez} 
Above $g^2/(\omega_0 t)=0.16$, the width of the pump pulse chosen to be $\tau_p=1/t$  
is comparable to the energy relaxation time and the relaxation occurs before turning off the pump pulse. 
In Figs.~\ref{fig:relax}(b) and (c), respectively, intensity maps of $O_i(\tau)$ in the space-time plane after the local excitation are presented for $\alpha=0.4$ and 0.8 at $\lambda=0.25$. 
It is shown that the wave fronts of the domain walls become broad and the averaged velocities decrease with time.  
This is clearly in contrast to the case of large $\alpha$ (see Fig.~\ref{fig:local_g}), in which the soliton  velocities are almost independent of time. 
The reduction of the soliton velocity is understood in the soliton model in Eq.~(\ref{eq:solitonmodel}): 
as shown in Eq.~(\ref{eq:velocity}), the soliton velocity is minimum (maximum) at the bottom (middle) in the soliton band.  
Thus, the energy relaxation from the fermionic to phononic systems leads to the reduction of the velocity, implying a suppression of the CO melting. 

 %---------------------------------------------------------------------------------------
\begin{figure}[t]
\includegraphics[width=0.9\columnwidth,clip]{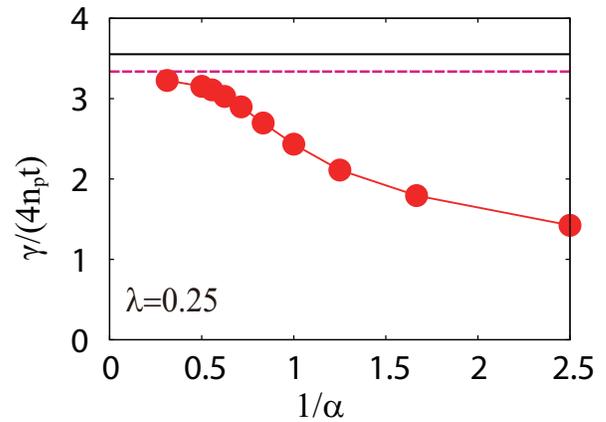}
\caption{
(Color online)
The normalized damping constant plotted as a function of $1/\alpha$ at $\lambda=0.25$. 
Black and red lines represent the results at $g=0$, and $g=0$ multiplied by 
$\exp^{-(g/\omega_0)^2}$. 
Other parameters are chosen to be $V/t=6$, $\omega_p/t=4$, $A_p=0.1$  and $\tau_p t=1$. 
}
\label{fig:damp_f}
\end{figure}
%---------------------------------------------------------------------------------------
%
Finally, characteristics in the damping factor is summarized; 
$\gamma/(4n_pt)$ are plotted as a function of the inverse of $\alpha$ in Fig.~\ref{fig:damp_f}. 
The damping factors  are obtained from the time profiles of $O(\tau)$ at a time when $v^\ast(\tau)$ is fully saturated.  
With increasing $\alpha$, the numerical data approaches to a red broken line, which represents the damping factor without the fermion-phonon coupling multiplied by the polaronic factor $e^{-(g/\omega)^2}$. 
Below $\alpha \sim 1$, reduction of the calculated results from the red line is remarkable and is attributable to the energy relaxation from the fermionic to phononic systems.

%\begin{figure}[t]
%\includegraphics[width=0.6\columnwidth,clip]{fig/band.eps}
%\caption{(Color online) Schematic illustration of the intraband transition of soliton.The solid curve represents the soliton band.The each circles represent the occupation of soliton band at each time,and the bold arrows represent the time evolution of soliton state.}
%\label{fig:band}
%\end{figure}

\section{Discussion and Summary}

We present in this paper a comprehensive analyses of the photoinduced melting dynamics of the one-dimensional CO state interacting with the local phonons. 
By using the iTEBD method, we obtain the reliable numerical data for the time dependences of the CO  parameter, the electronic and phononic energy and so on. 
We also analyze the CO melting dynamics in the system in which the artificial local photoexcitation is introduced. 
As well known, the excited state in the one-dimensional CO state without the electron-phonon coupling is described by the soliton and antisoliton picture. 
This is directly confirmed by the calculations of the local CO parameter induced by the local excitation shown in Fig.~\ref{fig:local_g0}; the domain-wall motions collapse the CO structure and give rise to the exponential time dependence of the CO parameter as $O(\tau)\propto e^{-\gamma \tau}$. 
The characteristic CO melting induced by the local excitation is reproduced by the classical soliton picture presented in Appendix A, in which the solitons are treated as classical particles. 
We have the expression for the damping constant $\gamma$ as shown in Eq.~(\ref{eq:classical}), being  proportional to the soliton velocity in the limit of the weak photoexcitation. 

The roles of the quantum phonons on the CO melting dynamics are characterized by the dimension-less adiabatic parameter $\alpha=\omega_0/t$. 
We perform the numerical analyses in a wide parameter region of $\alpha$.
%, and find that the transient dynamics show the crossover behavior from the adiabatic $(\alpha \lsim 1)$ to anti-adiabatic $(\alpha \gsim 1)$ regions. 
%
In the case of the antiadiabatic region ($\alpha \gsim 1$), the CO melting dynamics occurs by the polaronic soliton picture. 
The stable propagations of the domain walls are survived even by taking account of the fermion-phonon coupling, but the soliton velocity is renormalized. 
The damping constant $\gamma$ decreases with increasing the fermion-phonon coupling. 
This reduction of $\gamma$ in a wide region of the coupling constant ($0<g/\omega_0<1.25$) is well fitted by the expression for the polaronic state deduced by Lang and Firsov,~\cite{lang} which is known to be derived in the strong coupling regime. 
This implies that the solitons and ionic displacements propagate coherently even in the case of small $g$. 
This might be attributed to the fact that the ions are distorted cooperatively in the CO ground state even for small g, in which incoherent motions of electrons and ions produce a energy loss in macroscopic order. 
%
%to the local phonon in the one-dimensional lattice adopted in the present calculations; the local ion at the soliton site can distort without distortions at other sites. 

On the contrary, in the adiabatic case ($\alpha \lsim 1$), the CO melting in the early stage of the photoexcited transient state is not induced by the propagations of the solitions but their relaxations. 
The photogenerated solitons have a band energy of $\varepsilon_k=-2t \cos 2k$ and a velocity $v=4t\sin 2k$,  where the interaction with phonons and that between solitons are neglected. 
Trough the fermion-phonon coupling, the kinetic energy of the solitons are dissipated to the phononic system, and the soliton velocity is reduced. 
The solitons are settled down around the bottom of the band, at which these cannot be relaxed furthermore by emitting the quantum phonons. 
After the relaxation, the CO melting is induced by the soliton propagation, in the same way with the antiadiabatic case. 
This scenario is confirmed by the calculation of the transient electron energy and phonon energy, in which 
the relaxation time is scaled by $g^2/(\omega_0 t)$. 
The change of the time dependence of the CO parameter, that is, 
the non-exponential decay in the early stage and the exponential decay after the energy relaxation, is understood in this scenario.

We thank M.~Naka, H.~Seo and S.~Iwai for their helpful discussions.
This work was supported by MEXT KAKENHI Grant No. 26287070, 15H02100 and 17H02916. 
Some of the numerical calculations were performed using the facilities of the Supercomputer Center, the Institute for Solid State Physics, the University of Tokyo.

%%%%%%%%%%%%%%%%%%%%%%%%%%%%%%%%
%%%%%%%%%%%%%%%%%%%%%%%%%%%%%%%%
\appendix
\section{A classical soliton model}

\begin{figure}[t]
\includegraphics[width=0.8\columnwidth,clip]{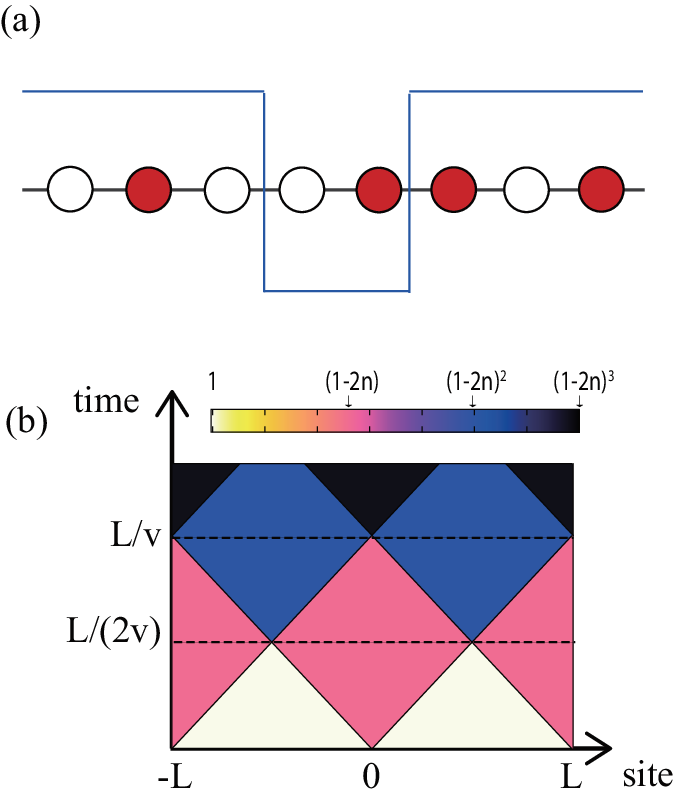}
\caption{
(Color online)
(a) A schematic picture for a soliton and an antisoliton in the CO state. 
(b) An intensity map of $O_i(\tau)$ in the space-time plane obtained by the classical soliton model. 
}
\label{fig:soliton}
\end{figure}

In this Appendix, we introduce a classical model for the photoinduced soliton and antisoliton in order to derive the analytical expression of the damping factor given in Eq.~(\ref{eq:classical}). 
We introduce a classical particle model in a one-dimensional chain, in which 
hard core particles (antiparticles) correspond to the solitons (anti-solitons) in the CO state. 
We assume that the (anti)particle velocity is constant $v$, and the elastic collision occurs between the (anti)particles. 
The pump photon irradiation is simulated by a stochastic generation of the particle-antiparticle pairs 
in the bonds connecting the nearest neighboring sites with a periodicity $L$ and probability $n$. 
We assume the $\pi$ phase shift of CO and a step-function like CO parameter across the (anti)soliton as shown in Fig.~\ref{fig:soliton}(a). 
The local CO parameter is calculated in this model and the intensity map of $O_i(\tau)$ in the space-time map 
is shown in Fig.~\ref{fig:soliton}(b). 
The result well reproduces the numerical results shown in Fig.~\ref{fig:local_g0}(a). 
At time $\tau_m=Lm/(2v)$ when the $m$-time collision occurs as indicated by broken lines in Fig.~\ref{fig:soliton}(b), the local order parameters are spatially uniform. 
We confirm at least up to $m=5$ that the amplitude of the CO parameter at the time $\tau_m$ are given approximately by $O(\tau_m)=(1-2n)^m$. 
This result is interpreted as follows. 
Let us consider the state at $\tau_1=L/(2v)$, until when the (anti)solitons propagate independently. 
The local CO parameters at the sites, where the (anti)soliton pass and does not pass over, are $1$ and $-1$, respectively. 
Since the probability for the photogeneration of solitons are $n$, the CO parameter at this time 
is obtained by the average calculation as  
$O(\tau_1)=1 \times (1-n) + (-1) \times n=1-2n$. 
At the time $\tau_2=2L/(2v)$, the second collision occurs between the soliton and antisoliton which are generated at $i=0$ and $L$. 
The four photoinduced states are possible: the (anti)soliton is generated at both the sites of $i=0$ and $L$, at either $i=0$ site or $i=L$ site, and neither the sites.  Since these probabilities are $n^2$, $n(1-n)$, $n(1-n)$ and $(1-n)^2$, respectively, the CO parameter is obtained by the average as $O(\tau_2)=1\times n^2 +(-1)\times n(1-n)+(-1) \times n(1-n) +1 \times (1-n)^2=(1-2n)^2$. 
The order parameters at $\tau_m$ with $m \geq 3$ are obtained in the same way.  
By using the relation $n=n_pL$ where $n_p$ is the effective photon number introduced in Sec.~\ref{sec:without_ep}, 
the CO parameter is given as $O(\tau_m)=(1-2n)^m=\exp[2v\tau_m \ln (1-2Ln_p)/L]$. 
This expression can be extended to the continuous time. 
By comparing this result with Eq.~(\ref{eq:exp}), the damping factor for the CO parameter is identified as  $\gamma=-2vL^{-1} \ln(1-2Ln_p)$ which is Eq.~(\ref{eq:classical}).

\end{document}